\documentclass[aps,pre,twocolumn]{revtex4-1}

\usepackage{amsmath}  
\usepackage{amsfonts} 
\usepackage{graphicx} 
\usepackage[english]{babel}
\usepackage[dvipsnames]{xcolor}
\usepackage{xfrac} 
\usepackage[normalem]{ulem}

\newcommand{\ud}{\,\mathrm{d}}

\newcommand{\ve}[1]{\mathbf{#1}} 

\newcommand{\vehr}{\hat{\ve{r}}}
\newcommand{\vehs}{\hat{\ve{s}}}
\newcommand{\brac}[1]{\left(#1\right)}

\newcommand{\Gammaf}[1]{\operatorname{\Gamma}\brac{#1}}

\newcommand{\pfrac}[2]{{#1}/{#2}}

\begin{document}

\title{Uniform line fillings}
\author{Evangelos Marakis}
\email{e.marakis@utwente.nl} 

\author{Matthias C. Velsink}

\author{Lars J. \surname{Corbijn van Willenswaard}}
\affiliation{Complex Photonic Systems (COPS),
MESA+ Institute for Nanotechnology,
University of Twente, PO Box 217, 7500AE Enschede, The Netherlands}

\author{Ravitej Uppu}
\altaffiliation{Present address: Center for Hybrid Quantum Networks (Hy-Q), Niels Bohr Institute, University of Copenhagen, Blegdamsvej 17, DK-2100 Copenhagen, Denmark}

\author{Pepijn W. H. Pinkse}
\affiliation{Complex Photonic Systems (COPS),
MESA+ Institute for Nanotechnology,
University of Twente, PO Box 217, 7500AE Enschede, The Netherlands}


\date{\today}

\begin{abstract}
Deterministic fabrication of random metamaterials requires filling of a space with randomly oriented and randomly positioned chords with an on-average homogenous density and orientation, which is a nontrivial task.
We describe a method to generate fillings with such chords, lines that run from edge to edge of the space, in any dimension. 
We prove that the method leads to random but on-average homogeneous and rotationally invariant fillings of circles, balls and arbitrary-dimensional hyperballs from which other shapes such as rectangles and cuboids can be cut. 
We briefly sketch the historic context of Bertrand's paradox and Jaynes' solution by the principle of maximum ignorance. 
We analyse the statistical properties of the produced fillings, mapping out the density profile and the line-length distribution and comparing them to analytic expressions. We study the characteristic dimensions of the space in between the chords by determining the largest enclosed circles and balls in this pore space, finding a lognormal distribution of the pore sizes.
We apply the algorithm to the direct-laser-writing fabrication design of optical multiple-scattering samples as three-dimensional cubes of random but homogeneously positioned and oriented chords.
\end{abstract}

\maketitle 

\section{Introduction} 

The physics of multiple light scattering in media is well-established \cite{Ish1978,Ros1999}. 
While random multiple-scattering media are abundant in nature, their microscopic structure originates from uncontrolled physical processes.
Recent breakthroughs in nanofabrication enabled the construction of multiple-scattering media with designer microscopic features as well as macroscopic properties such as transport mean free path and scattering anisotropy. 
Examples of such complex systems include photonic crystals \cite{Vos2015}, quasicrystals and deterministic aperiodic structures \cite{Led2015} and hyper-uniform media \cite{Mul2014}.
An emerging application of random multiple-scattering media is their use as optical physical unclonable keys in cryptography \cite{Pappu2002a, Goorden2014c, Uppu2018}. 
The security of these cryptography protocols relies on the assumption of technological infeasibility in creating a high-fidelity copy of a key. 
Maximal complexity of light scattering occurs in thick isotropic multiple-scattering media lacking any symmetry or long-range order \cite{Skipetrov2011}, which make them the best candidates for keys.  
We began the investigation into the robustness of the unclonability assumption by attempting to create the best possible replicas of isotropic multiple-scattering media with predetermined geometry.
While spherical scatterers ease the design of isotropic scattering media, conventional methods inhibit the precise positioning of the scatterers in a three-dimensional volume \cite{Voi2007}. 
Significant advances in direct laser writing (DLW) methods have overcome this limitation and enable precise positioning (up to few nanometers) of submicron-sized features \cite{Deubel2004, Renner2013}. 
The fabrication of stable structures using DLW typically involves an interconnected network of line segments that uniformly fill a three-dimensional volume of up to mm$^3$ \cite{Fischer2013}. 
In this paper, we address the problem of uniform filling of a space with chords with minimal long-range spatial correlation that realizes a design for photonic nanostructures using DLW.

\begin{figure}[b!]
\centering
\includegraphics[width=0.4\textwidth]{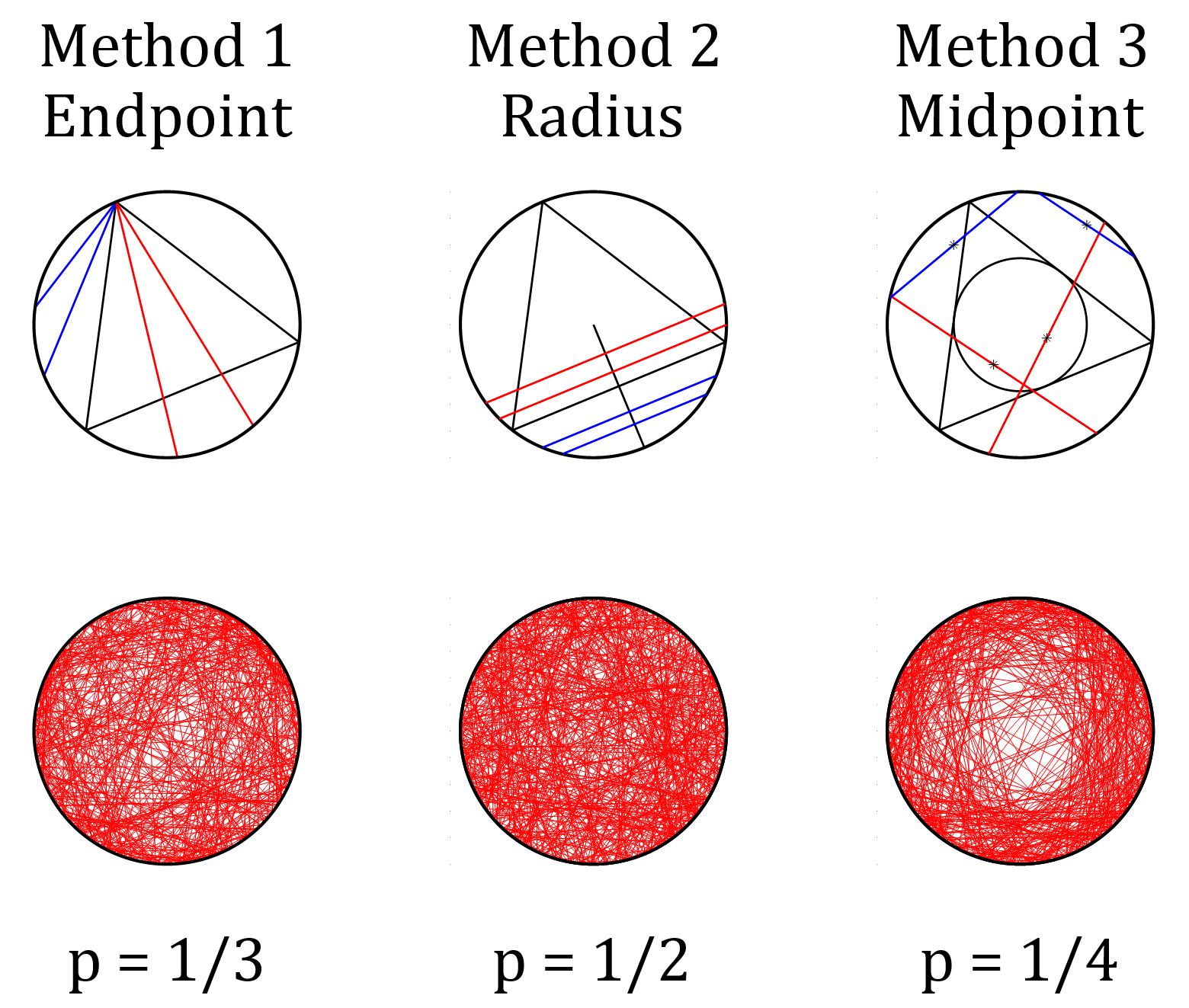}
\caption{Bertrand's paradox: The top row illustrates three ways of filling a circle with random lines: 1) connect two random points on the circumference of the circle,  2) choose a random radius and angle from the center, then draw the line perpendicular to that and, 3) choose a random point inside the circle and find the chord of which the point is the midpoint. The bottom row shows a single realization with 500 lines according to the method above it. The resulting probability $p$ of finding a chord longer than the edge length of the enclosed equilateral triangle is different for the three methods. Jaynes argued that only the “radius method” (middle column) results in ensemble-averaged random line filling that is translationally invariant, i.e. uniformly dense.  }
\label{case2d}
\end{figure}
The problem of filling space with randomly oriented chords is more involved than it seems at first glance. This was already noticed by J. Bertrand \cite{bertrand1889calcul}, who first formulated a paradox encountered with random line fillings in a circle: Suppose one fills a circle with random lines crossing the circle (such lines are also called ``chords'') and given the largest enclosed equilateral triangle, what is the probability $P$ that a random chord is longer than the side of the triangle? Perhaps surprisingly, the answer depends on the choice of the method to draw the random lines. Bertrand discussed three possible methods as illustrated in Fig.~\ref{case2d}, which result in three different probabilities $P$, concluding that specified as ``random filling'' only, the problem is ill-defined.

This paper deals with the generalization of the method for generating on-average uniform line fillings of a bounded domain in an $n-$dimensional space ($n\geq2$). We derive analytic expressions for the chord length distributions and validate the chord generation method using Monte Carlo simulations in two and three dimensions. Further, we analyze the size distribution of voids formed in uniform line fillings, which could provide insight into the structural correlations,
finding a lognormal distribution of the pore sizes.
We employ the line generation method to create designs of disordered multiple-scattering media that can be immediately implemented using DLW methods.
We believe our results will also be useful for other problems where random but straight pathways are found, such as e.g., dosimetry or sampling of non-stationary areas or volumes.

\section{Monte Carlo Method}
Let us begin by having a closer inspection of Bertrand's paradox.
In the first method, two points are chosen at random on the circumference of the circle and a chord is constructed. This method results in the probability $p=\sfrac{1}{3}$ of the so-constructed chord to be longer than the side of the enclosed equilateral triangle.
In the second approach, the ``radius method'', a line from the center of the circle of radius $r$ and an angle $0\le\phi\le2\pi$ is chosen at random and a chord perpendicular to this radial line is constructed. Since the base of the enclosed equilateral triangle is at exactly half the radius of the circle, this results in $p=\sfrac{1}{2}$.
In the third method, a random point inside the circle is chosen and the chord for which this point is the midpoint is found. The area in which midpoints of lines longer than the triangle's base are found is exactly $p=\sfrac{1}{4}$. The apparent conclusion from Bertrand's paradox is that there is not one random filling but that one should specify precisely what one means by a “random filling”. Jaynes argued that this is actually not necessary because of the principle of maximum ignorance \cite{Jaynes1973}, dictating a uniform random filling: If nothing has been specified on the specific location of the object in space, one has to assume that it does not matter, which is only the case for an on-average homogeneous filling. It should be mentioned that there is dispute in the literature on how complete Jayne's solution is \cite{rowbottom13,Drory15,Shackel07}. In this work we do not attempt to resolve this dispute but merely remark that Jayne's solution seems to be the most relevant for physical problems such as ours \cite{Liu18}.
This is  supported by more-recent literature on chord-length distributions that result when convex bodies are randomly intercepted by
straight lines such as in applications in acoustics, microscopy, texture analysis, and dosimetry \cite{Kel1984, Bor1994}.
In these studies the on-average homogeneous filling is also known as ``$\mu$-randomness'', in contrast to other random distributions such as ``$\nu$-randomness'' (straight lines through a random point in a sphere), ``$\lambda$-randomness'' (straight lines through two random points in a sphere), and ``i-randomness'' (ray originating in random point) \cite{Kel1984}.

Bertrand and Jaynes discussed line fillings of circles, but similar issues arise when filling other objects with lines. Such fillings can of course always be obtained by taking a larger circle and cutting out the lines intersecting with the smaller object, but one may wonder if there is a simpler method. For applications the distribution of line lengths in such objects quickly becomes non trivial and dependent on the geometry of the object \cite{Bailey2007,Philip2008}.
%

\begin{figure}[hbt!]
\centering
\includegraphics[width=0.5\textwidth]{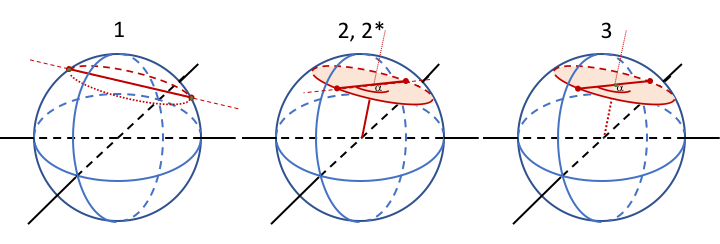}
\caption{Three-dimensional extension of the three methods of Bertrand. In 1) a line is drawn between two randomly-chosen points at the surface of the sphere. In 2, 2*) a randomly-oriented radial line is chosen and at a random radial distance $r$, a perpendicular line in arbitrary direction is selected. For method 2, $r$ is chosen from a uniform distribution $[0, 1]$, while for method 2*, $r^2$ is chosen uniformly from $[0,1]$. In 3) a random point in the ball is chosen and then a random line with that point as midpoint is chosen. In 3D, both method 1 and 2* give an on-average uniform distribution of lines. }
\label{case3d}
\end{figure}
\begin{figure}[hbt!]
\centering
\includegraphics[width=0.5\textwidth]{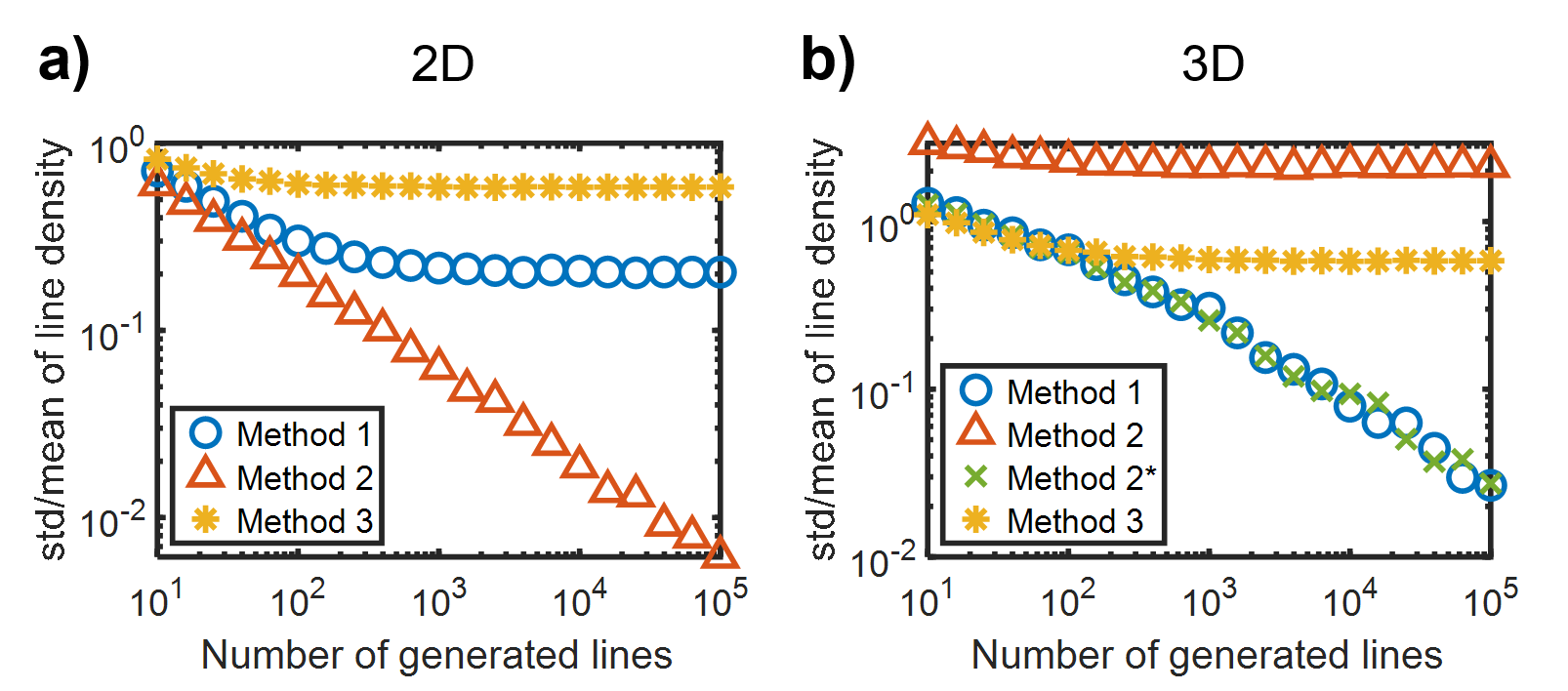}
\caption{Standard deviation over the mean of the line density for a) 2D and b) 3D. For 2D, Bertrand\textsc{\char13}s method 2 shows convergence, while both method 1 and 2* show convergence for 3D. }
\label{uniformityplots}
\end{figure}

It is straightforward to apply the three methods outlined in Fig.~\ref{case2d} in a Monte Carlo method to fill a circle with a random but on-average uniform distribution. In 3D, as illustrated in Fig.~\ref{case3d} the methods are extended to the filling of a ball. Method 1 in 3D connects two randomly chosen points on the surface of the sphere. In contrast to the 2D case, in 3D this method 1 produces an on-average uniform distribution. For the 3D version of method 2, a random directional distance $r<1$ is chosen from the center of the ball of unit radius. A randomly oriented direction perpendicular to this line segment is chosen to uniformly fill the ball. If $r$ is chosen from a uniform distribution $[0, 1]$, then the method does not lead to a homogeneously filling. If however, for a unit ball the radius is chosen according to the probability density function $f_R(r)=2r$, we obtain an on-average uniform filling. We call this method 2*. Method 3 in 3D selects a random point in the ball, finds the disc of those lines that have this point as midpoint and randomly picks a line from this disc. Interestingly, whereas in 2D only one method (2) produces a uniform distribution, in 3D we find there are two methods, 1 and 2*. In 3D we also found an additional method for generating random line fillings~\cite{Maz2004}.
\\
To check the three different methods by Bertrand for uniformity in a rigorous way, we use a Monte Carlo method implemented in Matlab to generate lines and test the convergence of the density. To test the convergence in 2D (3D), the circle (sphere) is divided into 100 shells of equal area (volume). The line density in that shell is then defined as the total line length inside it, divided by the shell\textsc{\char13}s volume. Figure~\ref{uniformityplots} shows the convergence of the line density for both 2D and 3D.

In order to fill other 2D shapes with lines, we will simply cut out such shapes from a larger circle. For the example of a unit square, we will produce a large number of lines with the random radius method in a disc of diameter $\sqrt{2}$. Then we cut out a square that is filled with random lines forming a homogeneous filling of space. The result is a homogeneous filling of the square of which a sample can be seen in the insert in Fig.~\ref{dist2d}.

\section{Characteristic Distributions}
\begin{figure}[hbt]
\centering
\includegraphics[width=0.5\textwidth]{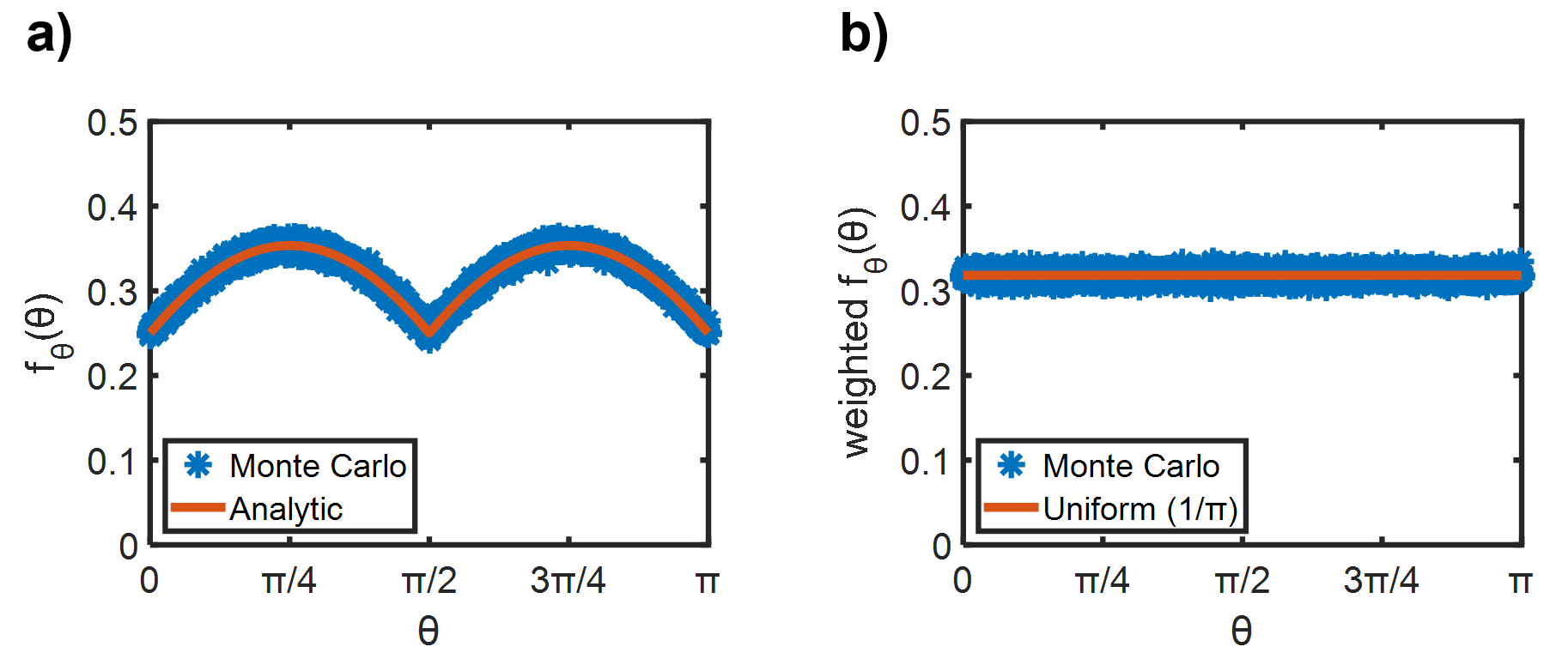}
\caption{a) Probability density function of the angles of $10^7$ chords in a square produced with our Monte Carlo code following the “random radius” recipe (method 2), counting the contribution of each line once, together with the function $ \frac{1}{4} (|\sin \theta |+| \cos⁡ \theta | ) $ (red line). b) The angular distribution weighing the lines with a factor proportional to their length. The result is a completely flat probability density function proving the angular homogeneity of the generated line filling.   }
\label{angles}
\end{figure}

Now that we have methods to generate random lines with an ensemble-averaged homogeneous distribution, it is helpful to look at characteristic distributions to compare the Monte-Carlo results with.
As a first test, we check the average angular distribution of the lines.
It is plotted in Fig.~\ref{angles}a).
We see that angles close to $\theta = 0$, $\pfrac{\pi}{2}$, $\pi$ (horizontal or vertical) are less likely whereas lines oriented along the diagonal directions $( \theta = \pfrac{\pi}{4}, \pfrac{3\pi}{4})$  are more likely. This represents the bias on the distribution caused by the angle-dependent cross section of the square for incoming parallel lines.
If we weigh the contribution of each of the lines with its length, we obtain the angular distribution plotted in Fig.~\ref{angles}b). Since the average line length ($\approx \frac{1}{2}\sqrt{2}$) for chords with $\theta \approx \pfrac{\pi}{4}, \pfrac{3\pi}{4}$ is smaller than the average line length for near-horizontal or near-vertical lines ($\approx 1$), the structure from  Fig.~\ref{angles}a) is flattened out. The resulting angular distribution is completely flat, representing the fact that in a homogeneously filled square the total line length for a certain angular range is constant.
%

\begin{figure}[t!]
\centering
\includegraphics[width=0.5\textwidth]{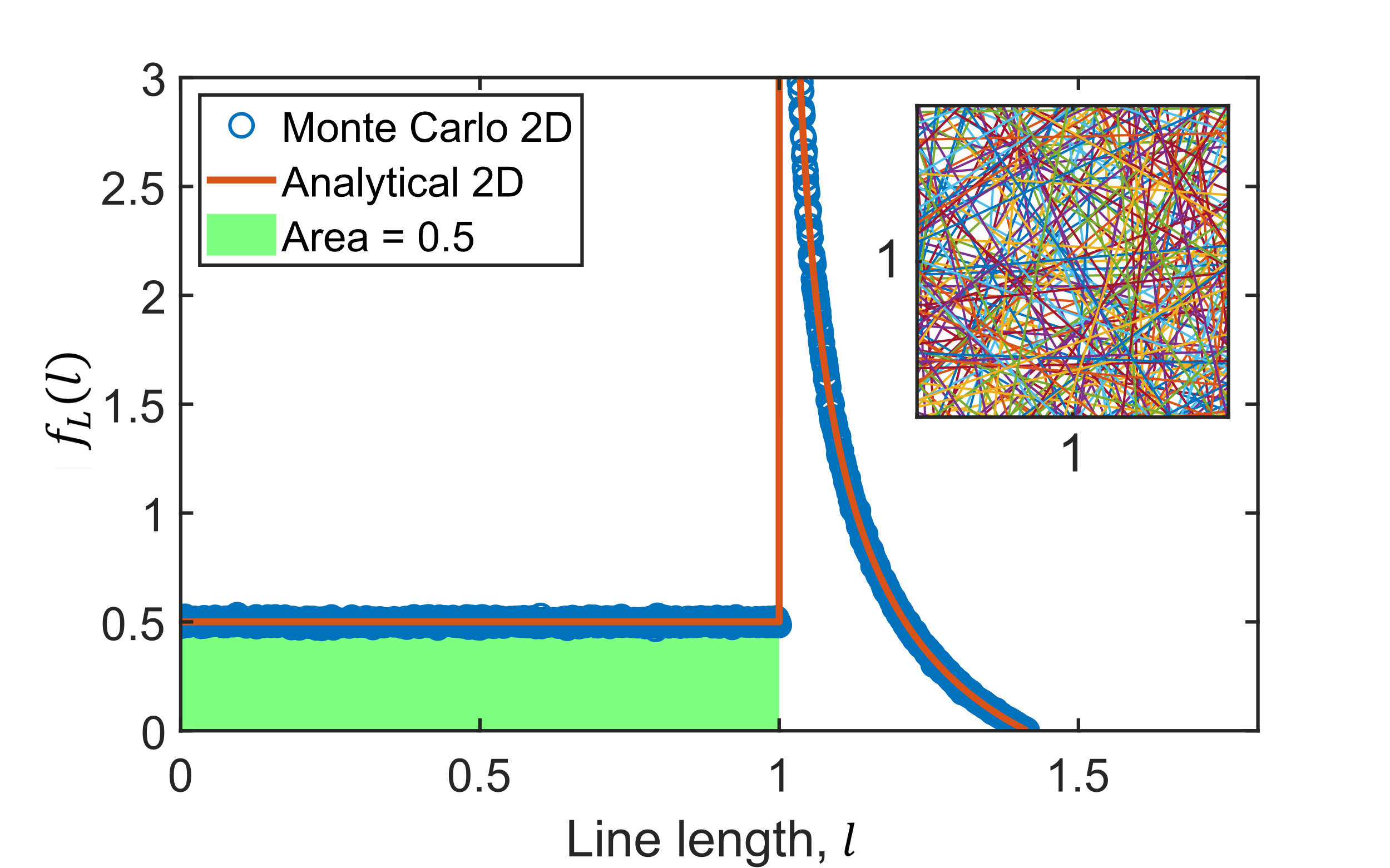}
\caption{ Length distribution of a homogeneous distribution of random lines in a unit square. The symbols are the result of our Monte Carlo method, the solid red line is the analytic result derived in the appendix. The longest possible line has a length of $\sqrt{2}$, but has a vanishing probability. The integrable singularity at $l=1$ is caused by the many possible lines that run almost parallel to one of the sides and therefore have a length of slightly more, but never less than one. Inset: a set of randomly chosen lines in a unit square.}
\label{dist2d}
\end{figure}

One of the more interesting distributions is that of line lengths.
The length distribution for the 2D lines is derived in the appendix and has a singularity at $l=1$, which is caused by the many possible lines that run almost parallel to one of the sides and therefore have a length of slightly more, but never less than one.
The distribution is plotted in Fig.~\ref{dist2d}, together with the statistical results according to the method discussed above with $10^7$ lines.
Interestingly, exactly half of the lines are shorter (or longer) than the side length. We suspect that there must be a simple geometric argument why this is the case, but have not found it.

\begin{figure}[hbt]
\centering
\includegraphics[width=0.5\textwidth]{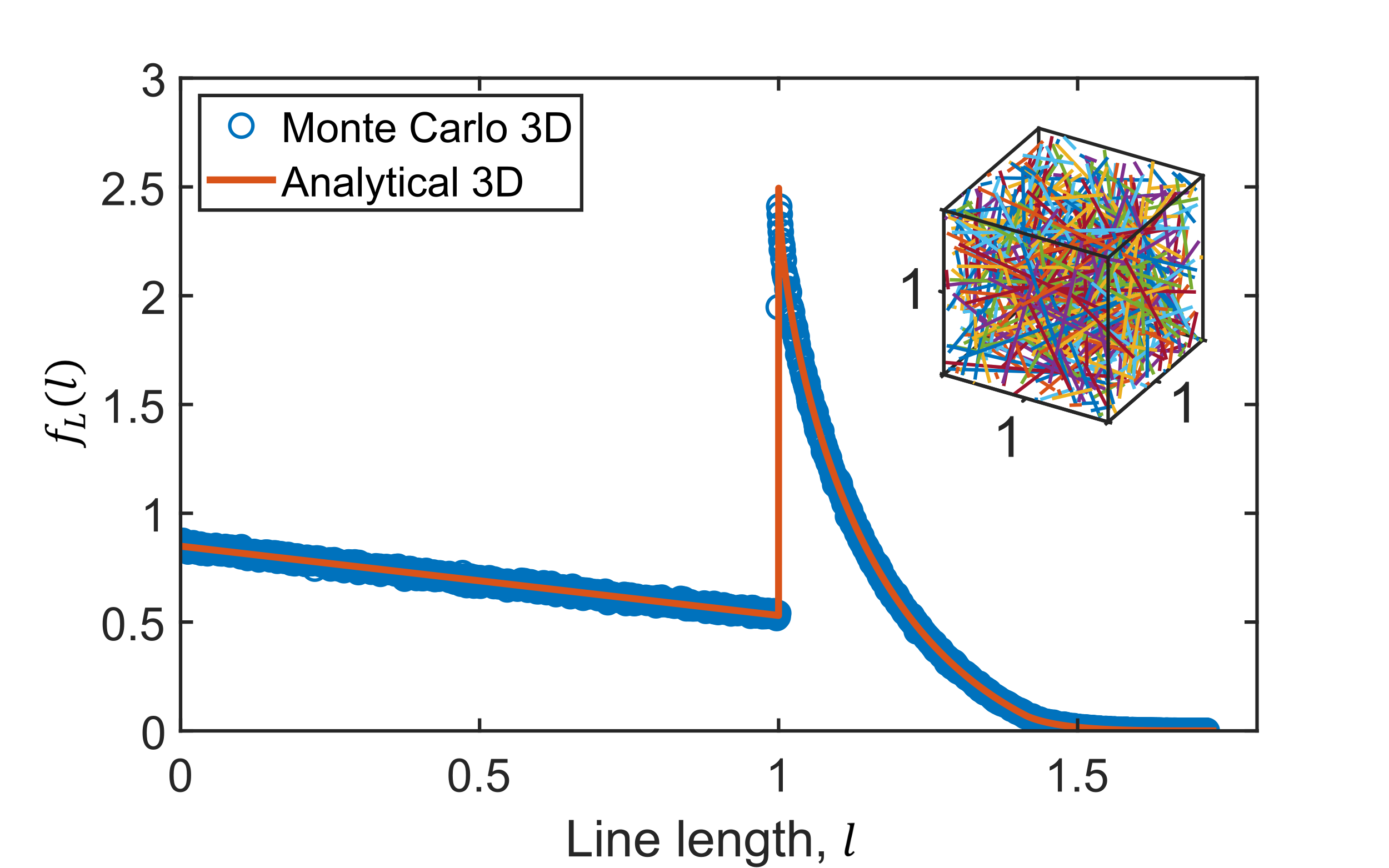}
\caption{ Length distribution of a homogeneous-on-average set of random lines in a unit cube. The symbols are the result of our Monte Carlo method, the solid red line is the analytic result given in the appendix. Inset: a set of randomly chosen lines in a unit cube.}
\label{dist3d}
\end{figure}
Using method 2* to create a uniform line filling of a cube, we obtain the distribution shown in Fig.~\ref{dist3d}. It resembles the distribution of lines in a square: the distribution has a singularity at $l=1$ and for shorter lines it is a straight line. However, the straight line for $0<l<1$ is not flat as in the case of the 2D square, but inclined: shorter lengths are more probable. The maximum possible length is now $\sqrt{3}$, and there is a kink at $\sqrt{2}$ expressing the fact that lines with a length between $\sqrt{2}$ and $\sqrt{3}$ are much less likely than lines shorter than $\sqrt{2}$, because these line lengths are only found with lines running close to one of the body diagonals of the cube. An analytic solution to the line length distribution of the cube was derived by Coleman \cite{Col1969} and is provided in the appendix.

\section{The void space}
We now turn to the voids between the homogeneously distributed chords. These voids are much more localized objects than the chords and have a characteristic size distribution that is important in a number of applications.
For instance, the light scattering properties of a sample of chords is more easily described by the properties of the set of voids than that of the lines.
In 2D, the voids between the lines can uniquely be determined by the polygons formed by the line crossings. This problem is equivalent to finding the ``pole of inaccessibility'' of a landmass, for which efficient algorithms exist if the landmass is convex \cite{martinez2012}. However, this method does not easily extrapolate to 3D, as the distance from a point to a line in 3D cannot be written as a linear function of the three coordinates of the point.

\begin{figure}[bt]
\centering
\includegraphics[width=0.4\textwidth]{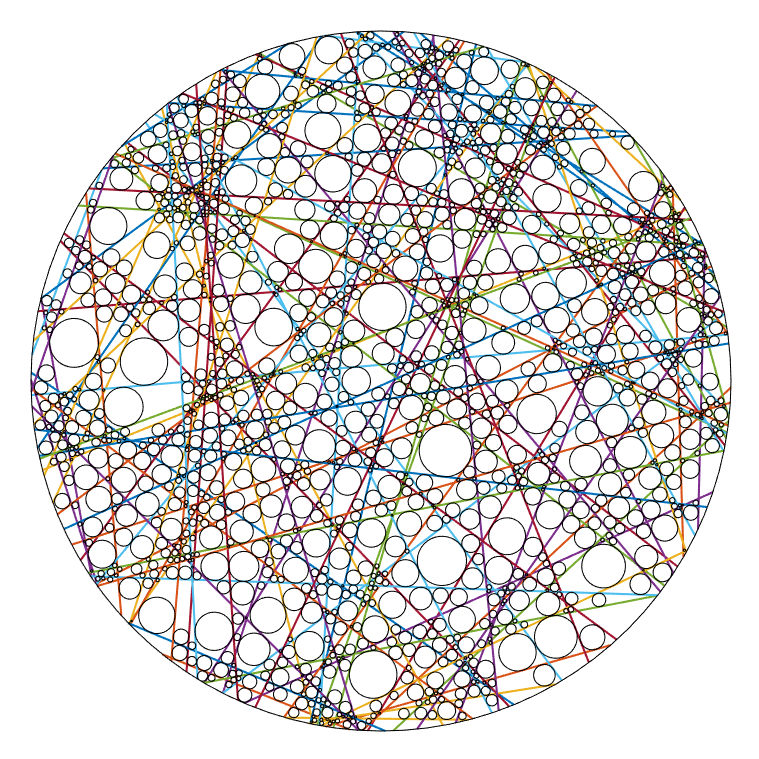}
\caption{Characterization of the void space by drawing the largest circles in the voids enclosed by the chords. In this sample distribution of 100 chords in a circle with unit radius, 1859 voids are identified. The line colors have no meaning and only serve to increase visibility. Inscribed circles cutting the large outer circle are discarded to avoid edge effects.}
\label{2DVoids}
\end{figure}

A method that does work in 3D is to find the largest spheres that are confined to the space between the lines \cite{Suf2015}. We find the largest inscribed sphere around a uniformly randomly picked starting point by maximizing the sphere radius with nonlinear optimization by gradient descent. Initially, the sphere is limited in size by the closest line. The gradient descent algorithm iteratively moves the sphere away from the closest line, until the sphere is bound by four lines (three lines in 2D) and cannot be further increased in size. Hence, the largest inscribed sphere in this region is found. Then we start the search again at another starting point. If a search enters the volume of an already maximum-sized sphere we discard that point and move to the next. To avoid the effect of the edge of the spherical sample region in which the lines are generated, an inscribed sphere is discarded if it cuts the outer bounds of the sample region, i.e. the unit sphere. If no new void is found after 1000 generated starting points, the void finding algorithm terminates. An illustration for the voids from a 2D line filling is shown in Fig.~\ref{2DVoids}. In each void the largest enclosed circle is drawn. The resulting radii of the thus-found circles are found to be lognormally distributed, but are not plotted here.

\begin{figure}[hbt]
\centering
\includegraphics[width=0.4\textwidth]{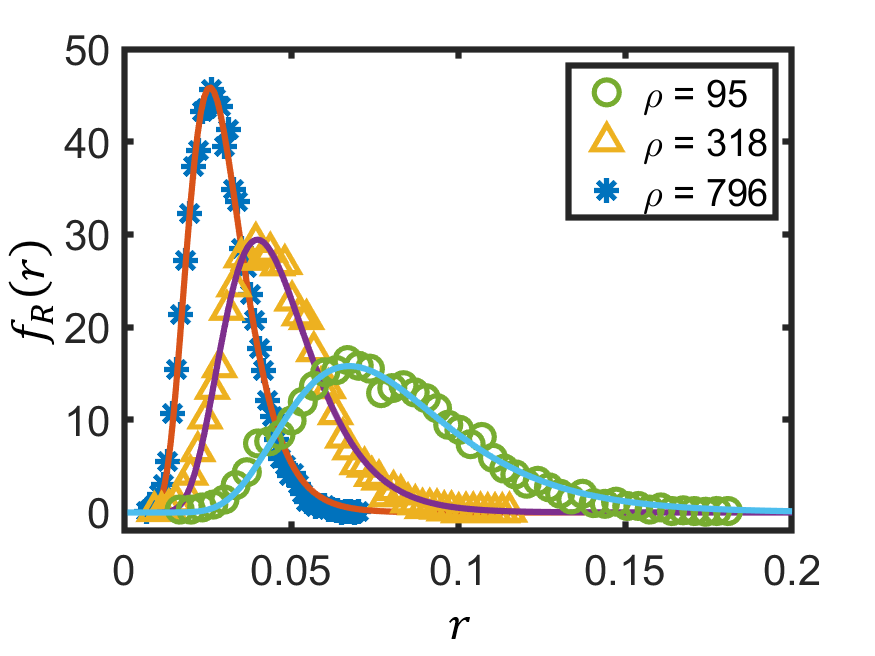}
\caption{Probability density function  of void radii in a 3D sphere of unit radius filled with a line density of $\rho=95$, 318, and 796 unit lengths per unit volume (symbols). The distributions are fitted with lognormal distributions (solid lines).}
\label{VoidSizeDistri}
\end{figure}
\begin{figure}[hbt]
\centering
\includegraphics[width=0.4\textwidth]{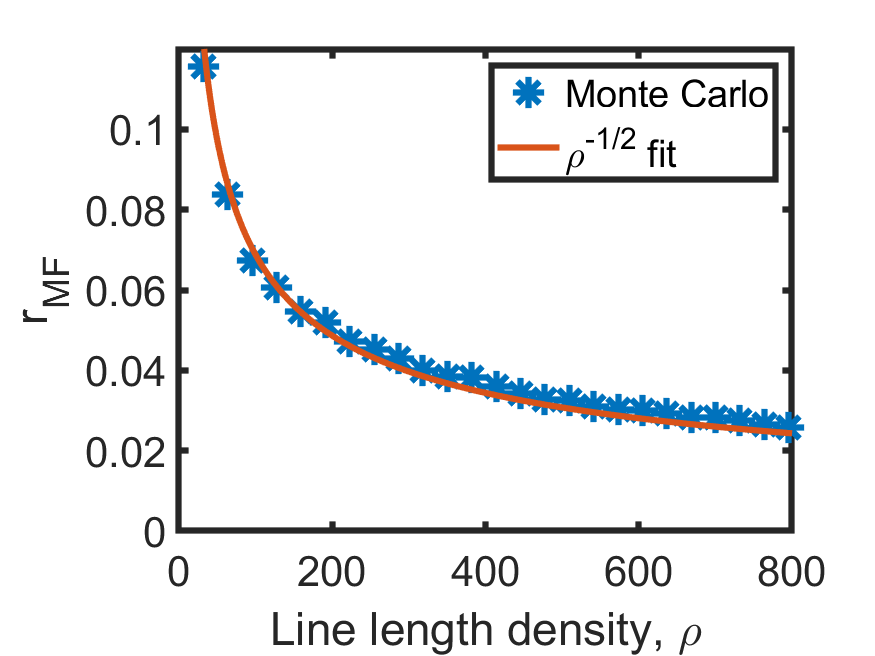}
\caption{The most frequent radius $r_{\rm MF}$ in the probability density function of void sizes in a 3D sphere filled with a line density $\rho$ in unit length per unit volume in combination with the predicted scaling behavior $1/\sqrt{\rho}$.}
\label{ScalingPlot}
\end{figure}

In contrast to the 2D case, the largest spheres that fit the voids in 3D can and will often overlap.  The probability density function ${f_R}(r)$ of the radii of these spheres is shown in Fig.~\ref{VoidSizeDistri} for three different line densities. Similar to 2D, the distribution of radii is well described by the lognormal distribution. 
The explanation is the following, where we start from a set of chords and its void size distribution. A new chord will cut a small fraction of the small voids as well as a few larger voids, effectively rescaling the size distribution a bit. Each additional chord will slightly rescale the distribution. All these many small rescalings become additive on a log scale, which leads via the central limit theorem to a normal distribution in log space.
The lognormal distribution is also found in physical distributions, such as nanoparticle agglomerates \cite{Den2016} or the voids between stars \cite{Rus2017}.
One might wonder how the void size distribution scales. For this we consider a $3$-dimensional sample cube of unit size with a single line. If we scale the size of this cube by a factor $m$, its volume will scale by $m^3$. The length of the line contained in this scaled cube will be a factor $m$ longer. For an arbitrary $d$-dimensional body containing multiple chords with lengths $L_i$, we can consider the line length density $\rho=\sum_i L_i/V$, with $V$ the volume of the body. Using the same arguments as for the cube, we can see that $\sum_i L_i$ scales as $m$, while $V$ scales as $m^d$. Hence the line length density will scale as $\rho\propto m^{1-d}$.  
Turning this argument around, a characteristic length scale of the voids is proportional to $ \rho^{1/(1-d)}$. Figure~\ref{ScalingPlot} shows this scaling in one example of such a characteristic length scale, namely the mode (the ``most frequent'') of the distribution of void sizes as a function of the density $\rho$. The deviation between the Monte Carlo data and the fit is caused by a small bias in the search algorithm that we employ to identify the voids: smaller voids are more likely to be missed by the randomly chosen starting points.

The following example on DLW-based structures helps in understanding the significance of this scaling law. Let us consider a $(20\,\mu{\rm m})^3$ cube with 400 homogeneously distributed random lines. They have (see appendix) an average length of $2/3$ units leading to a line length density of $400\times 2/3 =267$ unit length per unit volume, or $0.67\,{\rm \mu m}^{-2}$ in real-world units. Then the most frequent void size is $\approx0.045$ in units of the cube side length, which gives $0.045\times 20\,\mu{\rm m} =0.9\,\mu$m. 
This value of the most frequent void size holds for infinitesimally thin lines, which is physically impossible in nanofabrication. Assuming cylindrical lines with a finite thickness of $2t$, we can use another scaling law to estimate the effect: A uniform but non-zero value of $t$ reduces the radii of the inscribed spheres by $t$, thereby shifting the void size distribution down by $t$.
Mathematically we can describe this as the probability density function of the void sizes for thick lines, $f_R(r; t)$, which can be expressed in the probability density function of the zero-thickness lines as shown in Fig.~\ref{VoidSizeDistri} in the following way:  
$f_R(r; t) = c\, f_R(r+t)$ for all $r \geq 0$ and with $c$ a renormalisation factor. 
For instance, thick lines with a radius of $300\,$nm will in our example shift the most frequent void size down from $0.9\,\mu$m to $0.6\,\mu$m.

\section{$d$-Dimensional line filling}
We have shown how we can use the uniform line filling of a circle or 3D ball to uniformly fill a square or cube with lines. This can in fact be extended to arbitrary shapes in any-dimensional space as long as it can be embedded in a $d$-dimensional ball. For this we need to show two things, namely that we can extend method 2* to fill a $d$-dimensional unit ball and that this method gives a uniform line filling.

Extension of method 2* to $d$ dimensions is straightforward. To create a random line on a $d$-dimensional ($d \geq 2$) unit ball $B$, we use the following method:
\begin{enumerate}
    \item Choose a random direction vector $\vehr$ in $d$ dimensions.
    \item Select a random distance $r$ from the center using the probability density function $f_R^d(r)$. Let $\ve{p}$ be the point $r \vehr$.
    \item Select a random direction $\vehs$ in $d$ dimensions such that $\vehs \cdot \vehr = 0$ and construct the infinite line $\ve{p} + s \vehs$ for $s \in \mathbb{R}$.
    \item Take the intersection between this infinite line and the ball $B$ as the generated line.
\end{enumerate}

Since we choose $\vehr$ uniformly, there is no expected dominant angle in the lines. However, the density of the lines still depends on the distribution of the radii $r$ that we choose using $f_R^d(r)$. For $d=2$ we know that $f_R^2(r) = 1$ for a uniform density. Our Monte Carlo calculations in 3D gave a uniform density when $r^2$ is uniformly distributed, which corresponds to $f_R^3(r) = 2r$. We thus expect that using $f_R^d(r) = (d-1) r^{d-2}$ results in a uniform line density when filling the $d$-dimensional unit ball.

To show that this is the correct choice for $f^d_R(r)$, we look at a small $d$-dimensional ball $B(t)$ with radius $t < 1$ and volume $V_d (t)$, centered at the same origin as the original ball. Given a line generated by our method, we define its line length density on $B(t)$ as
\begin{equation}
    \rho(t) = \frac{l(t)}{V_d(t)},
\end{equation}
where $l(t)$ is the length of intersection between the generated line and $B(t)$. From Jaynes \cite{Jaynes1973} we know that the expected value of this quantity $\rho (t)$ should be independent of $t$ for uniformly distributed lines.

A line generated by our method can be characterized by three parameters ($\vehr$, $r$ and $\vehs$), each chosen randomly. Both $\vehr$ and $\vehs$ have effect on the orientation of the line, but not on the intersection length. The intersection length is thus determined only by $r$ and given by
\begin{equation}
    l(t) =
\begin{cases}
    0 & \text{if } r > t, \\
    2\sqrt{t^2 - r^2} & \text{if } r \leq t.
\end{cases}
\end{equation}
Since we choose $r$ probabilistically, we get a probabilistic line length $L(t)$. Its expected value is given by
\begin{equation}
    E[L(t)] = \int_0^{t} 2 \sqrt{t^2 - r^2} f_R^d(r) \ud r.
\end{equation}
The resulting expected line length density,
\begin{equation}
    E[\rho(t)] = \frac{E[L(t)]}{V_d(t)},
\end{equation}
should be constant for a uniform filling. To derive the constraint on $f_R^d(r)$ for such a constant line length density, we assume that the expected line length density $E[\rho(t)]$ has the unknown value $\overline{\rho} > 0$. Rewriting $E[\rho(t)] = \overline{\rho}$ gives
\begin{equation}
    \overline{\rho} V_d(t) = \overline{\rho} \frac{\pi^\frac{d}{2}}{\Gammaf{\frac{d}{2} + 1}} t^d
        = 2\int_0^t \sqrt{t^2 -r^2} f_R^d(r) \ud r,
\end{equation}
with $\Gammaf{\cdot}$ the gamma function. The solution to this Volterra integral equation is given by
\begin{align}
    f_R^d(r) & = \frac{\overline{\rho} \pi^{\frac{d}{2}}}{\Gammaf{\frac{d}{2}+1}}
    \frac{4r}{\pi} \brac{\frac{1}{2r} \frac{\ud}{\ud r}}^2
    \int_0^r \frac{t^{d+1}}{\sqrt{r^2 - t^2}} \ud t \nonumber\\
        & = \frac{\overline{\rho} \pi^{\frac{d}{2}}}{\Gammaf{\frac{d}{2}+1}}
    \frac{4r}{\pi} \brac{\frac{1}{2r} \frac{\ud}{\ud r}}^2
    \frac{\sqrt{\pi} r^{d+1} \Gammaf{\frac{d}{2}+1}}{2\Gammaf{\frac{d+3}{2}}} \nonumber\\
        & = \frac{(d+1)(d-1)}{2} \frac{\pi^{\frac{d-1}{2}}}{\Gammaf{\frac{d+3}{2}}}
        \overline{\rho} r^{d-2},
\end{align}
where we used that $d \geq 2$ and $r \geq 0$ to evaluate the integral. From this formula we see that $f_R^d(r)$ is of the form $a r^{d-2}$ for some constant $a$. However, we can not recover $a$ from this formula, since it depends on the unknown line length density $\overline{\rho}$. Therefore we use the fact that $f_R^d(r)$ is a probability density function and thus should be normalized. This results in $a = (d-1)$ and $f_R^d(r) = (d-1) r^{d-2}$, as expected. Thus the choice $f_R^d(r) = (d-1)r^{d-2}$ gives a uniform filling of a $d$-dimensional ball.

In addition to showing that this is the only correct choice for $f_R^d(r)$, we also find the average line length density of a single line
\begin{equation}
    \overline{\rho} = \Gammaf{\frac{d+1}{2}}\pi^{\frac{1-d}{2}}.
\end{equation}
This agrees with the expected value for the chord lengths \cite{Maz2003}. For applications with lines with small cross section $\sigma$, this line length density $\overline{\rho}$ can be used to compute the expected contribution of a single line to the fill fraction of the ball by $\overline{\rho} \sigma$. This also allows a rough estimate of the number of lines needed to fill an object to the desired fill fraction.

\section{Application}
\begin{figure}[hbt]
\centering
\includegraphics[width=0.5\textwidth]{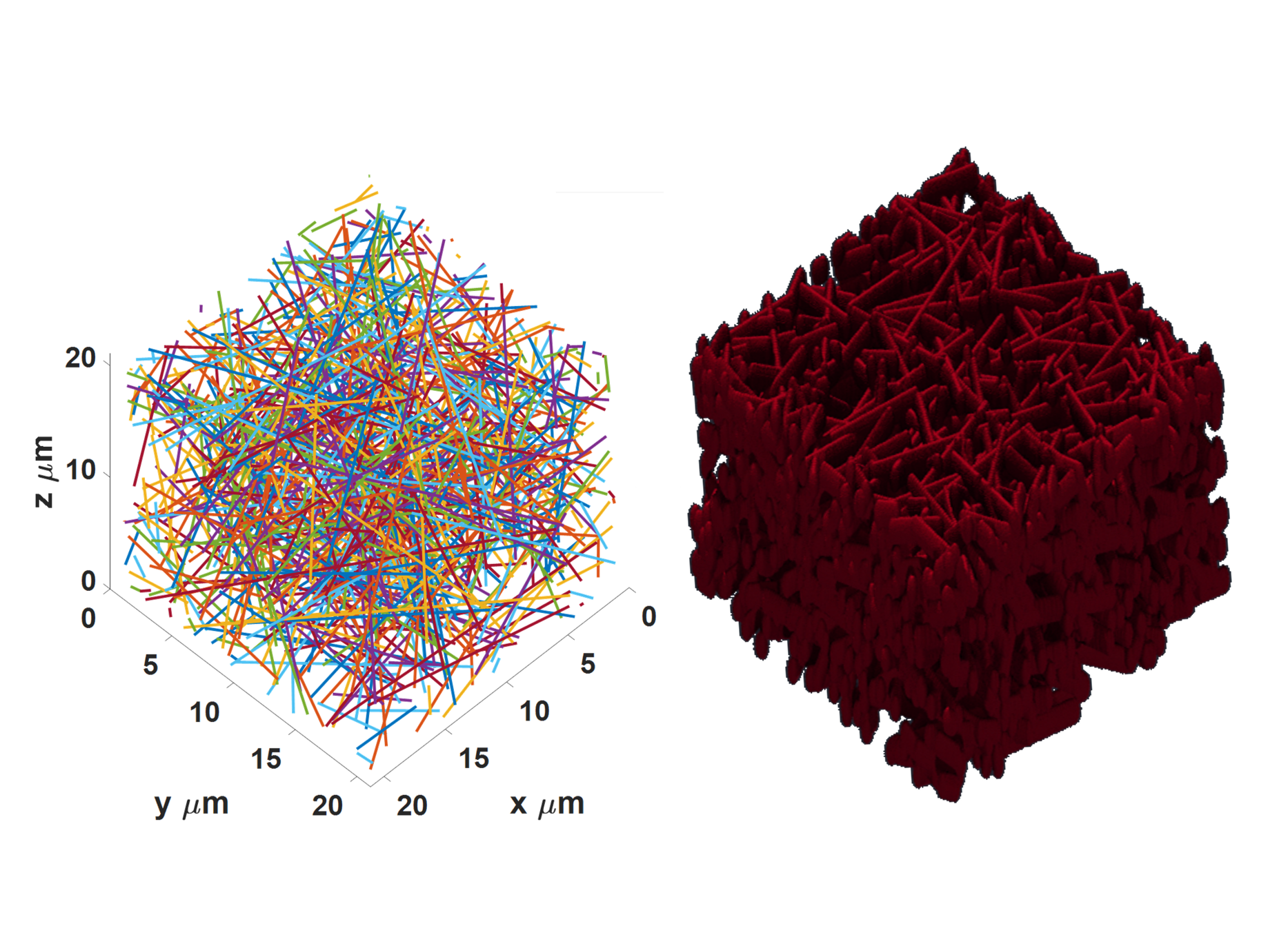}
\caption{ Line coordinates acquired with method 2* and fabrication design for DLW. The line coordinates of this methods can be used to create a deterministic scattering medium composed with polymer lines, ensuring uniformity and rotational translation invariant despite its finite volume (20 $\mu$m)\textsuperscript{3}}
\label{DLWSample}
\end{figure}

The motivation for our study was the wish to produce uniform-on-average random line fillings for the fabrication of deterministic scattering media. This would allow to investigate the assumption of unclonability made when using optical multiple-scattering media as physical keys for authentication \cite{Pappu2002a, Goorden2014c}.
Multiple-light-scattering media are composed of random inhomogeneities of the refractive index. Typical examples are fibers or powders of dielectric materials in air or in a matrix such as in paper or paint that give rise to speckle when illuminated with coherent light \cite{Goodman1976}. The nanoscopic geometry of the scatterers is uncontrolled. 
The nanofabrication method of our choice is direct laser writing (DLW), which combines - to our knowledge as the only 3D fabrication technique - an accuracy of few nanometers, submicron feature size, and a high index-of-refraction contrast with the ability to fabricate volumes up to mm$^3$ \cite{Fischer2013}.  
To take advantage of the deterministic fabrication of direct laser writing (DLW), we design small cuboid sample composed of randomly oriented rods.
With DLW it is straightforward to fabricate straight lines. At high enough density they make a strong interconnected network that creates a stable structure. 
Alternatively, the geometry of stacked spheres can be approximated by a mesh network \cite{Haberko2013}, but this remains a more complicated solution.
To ease comparison with radiative transport theory, on-average uniform and locally rotationally invariant samples are desirable. We used method 2* described above to design structures with on-average uniform density. We design cubes of deterministic scattering media with a length of 15 or 20 micrometer. Figure~\ref{DLWSample} shows a picture of the line coordinate map and the structure expected from the nanofabrication assuming a line thickness of $\approx 100\,$nm.

The optical investigation of these samples is the subject of future studies.
A deterministic multiple-scattering medium created with this method is expected to also be of interest for photovoltaics \cite{Burresi2013} and lighting industry \cite{Vos2013}. It has been shown that controlled scattering can improve the efficiency of solar cells by enhancing the dwell time of light inside the solar cell.
%
\section{Discussion and Conclusion}
Let us close with some remarks on interesting possible future directions. First of all, characterising the voids by finding the largest enclosed spheres is only one way to characterise the voids. 3D tessellation algorithms that also yield some shape information exist. One example is the Voronoi network that was also applied to a line filling \cite{Luc1999}. Once such tessellations are found, the chord length distribution of the pore space starts becoming non-trivial and can be studied such as is performed in the study of porous minerals \cite{Roz2007, Roz2014}. It is not clear to the authors, however, if such tessellations are uniquely defined. Given a tessellation, but also for our spherical delimiters, the pair correlation function and from that, the structure factor can be calculated. The latter is known to relate to the scattering mean free path \cite{Roj2004} and might, hence, be relevant for light scattering studies of actual optical scattering materials.

In summary, Jaynes\textsc{\char13} solution to Bertrand\textsc{\char13}s paradox, using the principle of maximum ignorance, is applied to generate homogeneous random line fillings of circles, spheres and hyperballs.  Arbitrary shapes with uniform random line fillings can be cut out of these geometries. We characterize the void space between the chords by finding the largest enclosed circles and 3D balls, finding they are lognormally distributed. We employ the random line fillings to design deterministic light scattering cubes of random dielectric rods that can be realized by direct laser writing. We believe our results will also be useful for statistical analysis, molecular dynamics and problems where random but straight pathways are beneficial, such as e.g., dosimetry, surveillance or sampling of non-stationary areas or volumes.

\section{Acknowledgements}
We thank Wouter Fokkema, Daan Frenkel, Matthias Schlottbom, Pim Venderbosch and Willem Vos for discussions. The project was partly financed by the Netherlands Organisation for Scientific Research (NWO).

\appendix{
\section{Analytic chord length distribution in a square and a cube}

\begin{figure}[hbt!]
\centering
\includegraphics[width=1.4in]{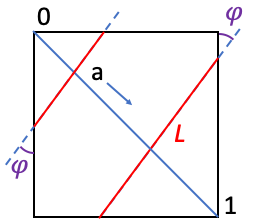}
\caption{ Construction to derive the length $L$
of a line segment (red) extending from edge to edge of a unit square (black) for a given angle $\phi$ $(0<\phi<\pfrac{\pi}{4})$ with the vertical and an intersection with the diagonal parametrized by $a$.}
\label{square}
\end{figure}
In this appendix we calculate, as an example, the distribution of chord lengths \cite{siegel_1978,ailam_1966} of the homogeneously filled 2D square. We start with finding an expression for the length of a line $L(a,\phi)$ (see Fig.~\ref{square}) as a function of the parametrized position where it crosses the diagonal, $a\in[0,1]$ as indicated in Fig.~\ref{square}, and the angle $\phi\in[0,\pfrac{\pi}{4})$ that the line makes with the vertical. Without loss of generality we can choose the angle to be in the domain $\phi\in[0,\pfrac{\pi}{4})$. All possible line lengths can be found in this domain (see Fig.~\ref{cdfplot} a).
\begin{equation}
L(a,\phi)=\\
\begin{cases}
\frac{1-a}{\sin \phi}+ \frac{1-a}{\cos \phi } & \text{if } a(1 + \tan{\phi}) \geq 1, \\
\frac{a}{\sin \phi}+ \frac{a}{\cos \phi } & \text{if } a (1 + \tan{\phi} )\leq \tan \phi, \\
\sec{\phi} & \text{if } a(1 + \tan{\phi}) <1\\
   \ \ \ & \;\& \; \tan{\phi} <a(1 + \tan{\phi}).
\end{cases}
\label{eq_L}
\end{equation}
%
\begin{figure}[hbt!]
\centering
\includegraphics[width=0.45\textwidth]{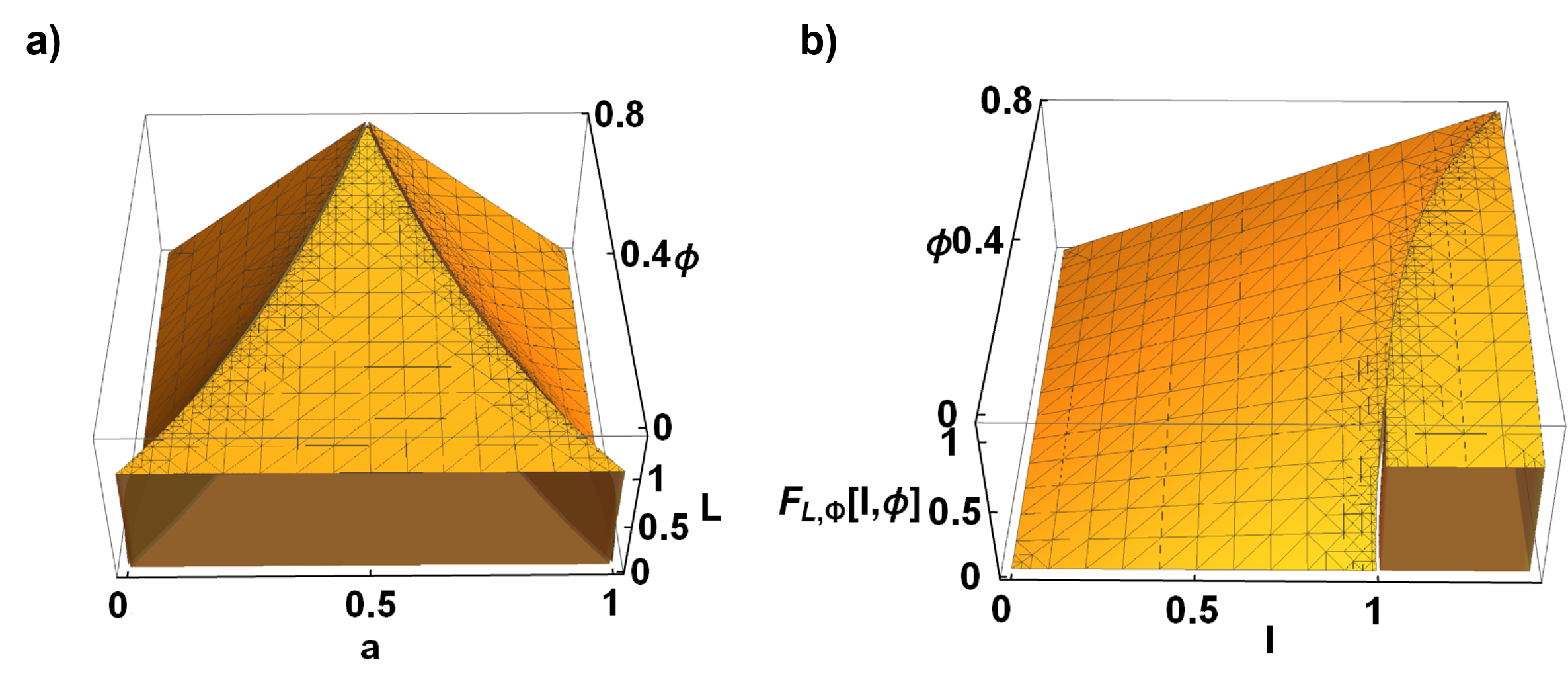}
\caption{a) The line length $l$ in a unit square as a function of the angle and parametrized entrance position, $a$. b) The cumulative density function, $F_{L,\Phi}(l,\phi)$, for the probabilities to find a line up to length $l$, still as a function of the angle $\phi$. Integration over $\phi$, taking into account the a priori probability to find a line with angle $\phi$, will yield the final $F_L(l)$.}
\label{cdfplot}
\end{figure}
The cumulative density function, $F_L(l)$, is found from this length function by integrating over the domains of $a$ and $\phi$ a function that is either $1$ or $0$ depending on the argument of the function, $l$, being larger or smaller than $L(a,\phi)$, respectively:
\begin{equation}
F_L(l)=\\
\frac{4}{\pi} \! \int\displaylimits_{\phi =0}^{\phi =\frac{\pi}{4}} \!\!\! P_\phi(\phi) \int\displaylimits_{a =0}^{a =1} \!\!\! P_a(a)
\left \{
  \begin{tabular}{ccc}
  1 &   & if $L( a, \phi ) \leq l$ \\
  0 &   & if $L( a, \phi ) > l$
  \end{tabular}
\right \}
\ud a \ud \phi ,
\end{equation}
\\
where $P_a (a)=1$ and $ P_\phi (\phi) =  \frac{1}{4} (|\sin \theta |+| \cos⁡ \theta | ) $ are the a priori probabilities for finding a line with position $a$ or angle $\phi$. Performing the integral yields:
\begin{equation}
F_L(l)=
  \begin{cases}
  \frac{l}{2} & \text{if }    0 \leq l < 1, \\
  1 - \frac{l}{2}+\frac{\sqrt[]{l^2 -1}}{l} & \text{if }  1 \leq l < \sqrt{2}.
  \end{cases}
\end{equation}
\\
From this the probability density function of the line length is derived by simply taking the derivative
\\
\begin{equation}
  f_L(l)=  \frac{\ud F_L(l) }{\ud l} =\begin{cases}
  \frac{1}{2} &  \text{if } 0 \leq l \leq 1, \\
  \frac{1}{l^2 \sqrt{l^2 -1}}-\frac{1}{l}    & \text{if } 1<l \leq \sqrt{2},
  \end{cases}
\end{equation}
\\
which is, of course, normalized as
\begin{equation}
\int_{0}^{\sqrt{2}} f_L (l) \ud l =1.
\end{equation}
The resulting chord length distribution $ f_L(l)$ was plotted in Fig.~\ref{dist2d}, together with the statistical results according to the Monte Carlo method discussed in the main text with $10^7$ lines.
The analytic result can readily be used to calculate the average chord length in a homogeneously filled unit square:
\begin{equation}
\int_{0}^{\sqrt{2}} l f_L (l) \ud l =\frac{\pi}{4},
\end{equation}
which is a useful quantity to estimate the average chord length of a homogeneously filled square with a known number of chords. It agrees with the more general expression for arbitrary convex 2D areas \cite{Col1969} for which the average chord length is given by $\pi A/P$, where $A$ is the surface area and $P$ the length of the perimeter.

For completeness we also reproduce the chord length distribution in a 3D unit cube as derived by Coleman \cite{Col1969}.
  \begin{align}
  & f_{L,{\rm3D}}(l) = \frac{1}{3\pi} \times \nonumber\\
  & \begin{cases}
  8-3l \! &  \text{if } 0 \leq l \leq 1, \\
  \frac{6\pi-1}{l^3}+6l-\frac{8}{l^3}(2l^2+1)\sqrt{l^2-1} \!    & \text{if } 1<l \leq \sqrt{2}, \\
  \frac{6\pi-5}{l^3}-3l+\frac{8}{l^3}(l^2+1)\\ \ \ \ -\frac{24}{l^3}\rm{arctan}(\sqrt{l^2-1})  \!\!\!  & \text{if } \sqrt{2}<l \leq \sqrt{3}
  \end{cases}
  \end{align}
%
The chord length distribution in cuboids and rectangular boxes has also been derived in earlier works \cite{Col1981}.
The analytic result for the line length distribution in the cube can be used to calculate the average line length in a homogeneously filled unit cube:
\begin{equation}
\int_{0}^{\sqrt{3}} l f_{L,{\rm3D}} (l) \ud l =\frac{2}{3},
\end{equation}
which is useful for instance in dosimetry problems. Again this agrees with the more general expression for arbitrary convex 3D volumes \cite{Col1969} for which the average chord length is given by $4V/S$, where $V$ is the volume and $S$ the surface area of the body.
Interestingly, the average chord length decreases when going from a square to a cube, although the maximum chord length increases from $\sqrt{2}$ to $\sqrt{3}$. Apparently it is more likely for a chord in 3D to "cut a corner" than in 2D.

Finally, for a 2D circle with unit radius, a homogeneous filling of chords leads to a chord length distribution of $f_{L,{\rm circle}}(l) =l/(2\sqrt{4-l^2})$ with an average chord length of $\pi/2$. For a 3D sphere with unit radius a homogeneous filling of chords leads to a chord length distribution of $f_{L,{\rm sphere}}(l) = l/2$ with an average chord length of $4/3$, both consistent with the more general result from \cite{Col1969}.
}

\end{document}